\documentclass[10pt]{iopart}
\usepackage{cite}

\renewcommand{\i}{{\rm i}}

\begin{document}

\title{Chaotic maps and flows: Exact Riemann-Siegel lookalike for
  spectral fluctuations}
\author{P. Braun$^{1,2}$ and F. Haake$^1$}
 \address{ $^{1}$Fachbereich Physik, Universit\"at Duisburg-Essen,
47048
 Duisburg, Germany}
\address{ $^2$Institute of Physics, Saint-Petersburg University,
198504 Saint-Petersburg,
  Russia}

\begin{abstract}
  To treat the spectral statistics of quantum maps and flows that are
  fully chaotic classically, we use the rigorous Riemann-Siegel
  lookalike available for the spectral determinant of unitary time
  evolution operators $F$. Concentrating on dynamics without time
  reversal invariance we get the exact two-point correlator of the
  spectral density for finite dimension $N$ of the matrix
  representative of $F$, as phenomenologically given by random matrix
  theory. In the limit $N\to\infty$ the correlator of the Gaussian
  unitary ensemble is recovered. Previously conjectured cancelations
  of contributions of pseudo-orbits with periods beyond half the
  Heisenberg time are shown to be implied by the Riemann-Siegel
  lookalike.
\end{abstract}

\maketitle

\section{Introduction}

Universal fluctuations in quantum energy spectra  under conditions
of full classical chaos are well understood in terms of
Gutzwiller's semiclassical periodic-orbit theory
\cite{Heusl07,Muell09}. The phenomenological description
previously given by random-matrix theory has been fully recovered
for individual (rather than ensemble averages of) systems from the
unitary, orthogonal \cite{Heusl07,Muell09}, and symplectic
\cite{Braun12} symmetry classes.

The analogous problem for periodically driven dynamics will be
taken up in the present paper. A unitary Floquet operator $F$
describing the evolution of such systems over one period of the
driving and its quasi-energy spectrum will be investigated.
Limiting ourselves to dynamics without time reversal invariance,
we shall aim at recovering the spectral fluctuations of the
circular unitary ensemble of random-matrix theory. Surprisingly,
the semiclassical analysis will not only produce the two-point
correlator of the density of quasi-energy levels in the limit of
infinite dimension $N$ of the Floquet matrix (which coincides with
the correlator of energy levels of autonomous dynamics), but even
the finite-dimension variant (a periodic function of the spacing
variable, with $N$ times the mean level spacing as the period).

Floquet operators turn out easier to analyze than Hamiltonians.
It is the so called selfinversiveness of the secular polynomial of
unitary matrices which comes to our help; it provides a variant of
the secular
polynomial, a ''zeta'' function %
$\zeta(\varphi)=\sqrt{\det(-F^\dagger)}\e^{-\i
  N\varphi/2}\det(1-\e^{\i\varphi}F)$, %
which is real for real values of the phase variable $\varphi$ and
enjoys a rigorous finite-$N$ Riemann-Siegel lookalike.  Four zetas
will be combined as $Z=\langle\zeta(\varphi_C) \zeta(\varphi_D)
\zeta(\varphi_A)^{-1} \zeta(\varphi_B)^{-1}\rangle$ to a
generating function $Z$ which latter has the two-point correlator
of interest as a suitable derivative. Due to the fixed order $N$
of $\det(1-\e^{\i\varphi}F)$, the generating function is
represented by finite Fourier series with respect to the variables
$\varphi_C,\varphi_D$; the Fourier coefficients are related to the
traces of powers of the Floquet operator, $t_n={\rm Tr}F^n$, by
the venerable relations given by Isaak Newton nearly 350 years
ago. In contrast, the inverse zetas in $Z$ have
infinite Fourier series where the Fourier coefficients can again
be expressed in terms of the traces $t_n$. Inasmuch as those $t_n$
are determined by a Gutzwiller trace formula we get semiclassical
access to $Z$; the trace formula for $t_n$ is a well behaved one,
without divergence necessitating regularization.

When scrutinizing the semiclassical limit of $Z$ according to the
strategy just sketched we shall meet with a remarkable property, long
hoped for but hitherto elusive. It is well to highlight that property
right away in intuitive semiclassical parlance, before pointing to the
rigorous algebraic origin. The finite Fourier series representing $Z$
with respect to the variables $\varphi_C,\varphi_D$ have the welcome
consequence, due to the Riemann-Siegel lookalike, that periodic orbits
with periods only up to half the Heisenberg time determine the
constituent zetas $\zeta(\varphi_C)$ and $\zeta(\varphi_D)$. On the
other hand, orbits with periods up to infinity enter the infinite
Fourier series of the inverse zetas in $Z$. That 'asymmetry' in the
pairs $\varphi_A,\varphi_B$ and $\varphi_C,\varphi_D$ makes for a
vexatious inconvenience in evaluating $Z$. In previous work on
autonomous flows the analogous difficulty was fought with the help of
imaginary parts of the energy variables large enough to cut off the
contributions of orbits with periods beyond half the Heisenberg
time. In our present case, no trick at all is needed. Orbits with
periods larger than half the Heisenberg time make exactly zero
contribution to $\zeta(\varphi_C)\zeta(\varphi_D)$ in $Z$. The
algebraic basis for the property under discussion is the rigorous
Riemann-Siegel lookalike for the zeta function. That powerful identity
forces the coefficients of the secular polynomial
$\det(1-\e^{\i\varphi}F)=\sum_{n=0}^N A_n\e^{\i n\varphi}$ with $n>N$
(and thus the coefficients of $\e^{\i n\varphi}$ in the finite Fourier
series for $\zeta(\varphi)$ with $|n|>\frac{N}{2}$) to vanish --- even
though Gutzwiller's periodic-orbit sum for the traces $t_n$ contains
no information of the finite dimension of the Floquet matrix.

The rigorous Riemann-Siegel lookalike can be carried over to the
treatment of the energy level fluctuations for autonomous flows. One
has to resort to a stroboscopic description and work with the Floquet
operator for a suitable strobe period which secures a one to one
relation between $N$ energy levels and eigenphases. Again, finite
Fourier series arise for the generating function $Z$ for the phases
$\varphi_C,\varphi_D$. For the unitary symmetry class, the diagonal
approximation yields, in the limit $N\to\infty$, the exact $Z$ known
from the Gaussian unitary ensemble. As regards corrections we show
that the present approach reproduces, for $N\to\infty$, the previous
results for all  Wigner-Dyson symmetry classes.

The rest of the paper is organized as follows. In Sect.~\ref{secdet}
we present the Riemann-Siegel lookalike for the zeta function of
finite dimensional unitary matrices $F$. Newton's relations are
employed to express zeta in terms of traces of powers of $F$. In
Sect.~\ref{generatingfunction} the generating function is defined and
decomposed in additive pieces according to the Riemann-Siegel
lookalike. In Sect.~\ref{semiclassics driven} we invoke the Gutzwiller
trace formula for powers of Floquet operators and show the diagonal
approximation for the unitary symmetry class to produce the exact
generating function known from Random Matrix Theory. Some hints are
given on why we are certain the vanishing of corrections can be shown
in a generalization of previous work on autonomous flows.  In
Sect.~\ref{mapsforflows} we argue that autonomous flows can be
described stroboscopically in terms of unitary Floquet operators. In
the final section~\ref{summary} we summarize the progress made through
the rigorous Riemann-Siegel lookalike, comment on related previous
work, and speculate on future ramifications.

\section{Riemann-Siegel lookalike for unitary matrices}\label{secdet}

 Towards capturing spectral fluctuations we look at the secular
determinant
\begin{equation}
\label{eq:1}A(\varphi)=\det(1-\mathrm{e}^{\i\varphi}F)=\sum_{n=0}^{N}%
\,A_{n}\mathrm{e}^{\i n\varphi}%
\end{equation}
where the coefficients $A_{n}$ include $A_{0}=1$ and $A_{N}=\det(-F)$.
The determinant $A(\varphi)$ is defined in the whole complex $\varphi$-plane.
We may think of the determinant as a discrete one-sided Fourier transform of
the coefficients with the inverse
\begin{equation}
\label{eq:2}A_{n}=\int_{0}^{2\pi}\frac{d\varphi}{2\pi}\,A(\varphi)
\mathrm{e}^{-\i n\varphi}\,.
\end{equation}

The information stored in $A(\varphi)$ is also encoded in
the (trace of the) resolvent,
\begin{equation}
\label{eq:16a}t(\varphi)=\mathrm{Tr\,}\frac{1}{1-\mathrm{e}^{{\i}\varphi}F}\,.
\end{equation}
To decode, a geometric series for the resolvent is useful, and that series
takes different forms in the upper and lower halves of the complex $\varphi
$-plane. Defining the 'traces' (of powers of the Floquet operator),
\begin{equation}
\label{eq:17}t_{n}=\mathrm{Tr\,} F^{n}\,, \qquad\qquad n=0,\pm1,\pm2, \ldots
\end{equation}
we have
\begin{equation}
\label{eq:3}t(\varphi)=\left\{
\begin{array}
[c]{ll}%
\;\;\sum_{n=0}^{\infty}\,\mathrm{e}^{\i n\varphi}\,t_{n} & \quad
\mathrm{Im}\,\varphi>0\\
\!\!\! -\sum_{n=1}^{\infty}\,\mathrm{e}^{-\i n\varphi}\,t_{n}^{*} &
\quad\mathrm{Im}\,\varphi<0
\end{array}
\right. \,.
\end{equation}
In both cases, imaginary parts of the phase may be seen as convergence
ensurers. As above for the secular coefficients, we recover the traces $t_{n}$
from the resolvent by inverting the one-sided Fourier transform, $t_{n}%
=\int_{0}^{2\pi}\frac{d\varphi}{2\pi}\,t(\varphi) \mathrm{e}^{-\i n\varphi}$.

Inasmuch as the determinant of any matrix equals the exponentiated trace of
its logarithm we can express the resolvent in terms of the determinant as
\begin{equation}
\fl \label{eq:4}t(\varphi)=N+{\i}\frac{\partial}{\partial\varphi}
\ln A(\varphi) \quad\leftrightarrow\quad
A(\varphi)=\exp\Big(-\i\int^{\varphi}d\varphi
^{\prime}\;\big(\,t(\varphi^{\prime})-N\big)\Big)\,.
\end{equation}
The foregoing relation leads us to different expressions for the secular
determinant for the two possible signs of ${\rm Im}\, \varphi$,
\begin{equation}
\label{eq:5}A(\varphi)=\left\{
\begin{array}
[c]{ll}
\,\exp\Big(-{\textstyle \sum_{n=1}^{\infty}\frac{t_{n}\mathrm{e}^{\i n\varphi
}}{n}}\Big) & \quad\mathrm{Im}\,\varphi>0\\
A_{N}\exp\Big(\i N\varphi-{\textstyle \sum_{n=1}^{\infty}\frac{t_{n}
^{*}\mathrm{e}^{-\i n\varphi}}{n}}\Big) & \quad\mathrm{Im}\,\varphi<0
\end{array}
\right. \,.
\end{equation}
We have fixed integration constants by invoking $A(\varphi)\to1$
for $\mathrm{Im}\,\varphi\to+\infty$ in the first case, and
$A(\varphi)\to A_{N}\mathrm{e}^{\i N\varphi}$ for
$\mathrm{Im}\,\varphi\to-\infty$ in the second case. By expanding
the right members of (\ref{eq:5}) in powers of
$\mathrm{e}^{\pm\i\varphi}$ we formally get infinite series.
However, the traces $t_{n},\,t_{n}^{*}$ conspire to secure
$A_{n}=0$ for $n>N$. Even on the real $\varphi$-axis the two
expressions exist and coincide, provided $\varphi$ does not equal
any of the eigenphases of $F$.

The traces $t_{n}$ and the secular coefficients $A_{n}$ are
directly related as well. The Taylor expansion of (\ref{eq:5})
gives $ A(\varphi)=\sum_{V=0}^\infty
\frac{(-1)^V}{V!}\left(\sum_{l=1}^\infty \frac{t_l}{l}\e^{\i
l\varphi}\right)^V $. The secular coefficients are obtained by
employing the multinomial expansion for the powers $(\cdot)^V$ and
collecting the coefficients of  $\e^{\i n\varphi}$ with
$n=0,1,\ldots$.  To formulate the result, we consider all
partitions of the positive integer $n$ into sums of smaller
positive integers $l$ with multiplicities $v_{l}$, namely $n=\sum
_{l=0,1,\ldots}lv_{l}$. For the 'vector' $\vec{v}$ we define two
properties, $L(\vec{v})=\sum_{l=0,1,\ldots}lv_{l}$ and
$V(\vec{v})=\sum_{l=0,1,\ldots }v_{l}$. We then have
\cite{Krish95,Kotto99,Haake10}),
\begin{equation}
\label{An}A_{n}=\sum_{\vec{v}}^{\{n=L(\vec{v})\}}(-1)^{V(\vec{v})}
\prod_{l\geq1}\frac{t_{l}^{v_{l}}}{l^{v_{l}}v_{l}!}\,.
\end{equation}
It is remarkable that the expression (\ref{An}) holds independently of
the degree $N$ of the secular polynomial. Nevertheless, if the traces
$t_{n} $ are evaluated for fixed $N$ we get $A_{n}=0$ for $n>N$. The
expression (\ref{An}) can be regarded as the explicit solution of
Isaak Newton's recursive relation
$(N-n)A_{n}=\sum_{m=0}^{n}t_{m}A_{n-m}$ for the secular coefficients;
for that reason we shall refer to it as the Newton relation.

The unitarity of the Floquet operator entails what is known as the
'self-inversiveness' of the secular coefficients \cite{Haake10},
\begin{equation}
\label{eq:7}A_{n}=A_{N-n}^{*}A_{N}\,.
\end{equation}
That property makes sure that the set of the secular coefficients
contains exactly $N$ independent real parameters (as which one may
see the eigenphases of $F$). For $N$ odd, the $A_{n}$ with
$n=1,2,\ldots, \frac{N-1}{2} $ together with $A_{N}$ fully
represent the secular polynomial since the remaining coefficients
are given by selfinversiveness according to (\ref{eq:7}).
Similarly, for $N$ even, the $A_{n}$ with $n=1,2,\ldots,
\frac{N}{2}$ carry the full information. As a further consequence
of selfinversiveness, the following variant of the secular
determinant, to be called 'zeta function',
\begin{eqnarray}
\label{zeta}\zeta(\varphi)  =\sqrt{A_{N}^{*}}\mathrm{e}^{-\i N\varphi
/2}A(\varphi)
\end{eqnarray}
is real valued for real $\varphi$. For the case of even $N$
which we shall stick to from
here on, it is convenient to split the zeta function in three additive pieces,
\begin{eqnarray}
\label{zetaeven}%
\fl \zeta(\varphi)  =\sqrt{A_{N}^{*}}\mathrm{e}^{-\i
\frac{N}{2}\varphi}\sum_{n=0}^{\frac{N}{2}-1} A_{n}\mathrm{e}^{\i
n\varphi} + \sqrt{A_{N}^{*}} A_\frac{N}{2} +
\sqrt{A_{N}}\mathrm{e}^{\i \frac{N}{2}\varphi}
\sum_{n=0}^{\frac{N}{2}-1} A_{n}^{*}\mathrm{e}^{-\i n\varphi}\nonumber\\
 \equiv\zeta_{-}(\varphi) +\zeta_0 +
 \zeta_{+}(\varphi)\,.
\end{eqnarray}
The terms $\zeta_-$ and $\zeta_+$ contain Fourier components $\e^{\i
  \nu\varphi}$ with, respectively, $\nu=-1,-2,\ldots, -\frac{N}{2}$
and $\nu=1,2,\ldots,\frac{N}{2}$ while $\zeta_0$ is the zeroth Fourier
component, $\zeta_0=\sqrt{A_{N}^{*}} A_\frac{N}{2}=\sqrt{A_\frac{N}{2}
  A_\frac{N}{2}^*}=\zeta_0^*$.  The decomposition (\ref{zetaeven}) of
zeta can be seen as a rigorous \textit{Riemann-Siegel lookalike}
\cite{Kotto99,Haake10,Berry90, Berry92,Keati92, Bogom92}.

An alternative form for the zeta function is useful. It is
obtained by importing the representation (\ref{eq:5}) of
$A(\varphi)$ to the definition (\ref{zeta}),
\begin{equation}
\label{altzeta}\zeta(\varphi)=\left\{
\begin{array}
[c]{ll}%
\sqrt{A_{N}^{*}}\exp\Big\{-\i \frac{N}{2}\varphi -\sum_{n=1}
^{\infty}\frac{t_{n}\mathrm{e}^{\i n\varphi}}{n}\Big\} & \quad\mathrm{Im}
\,\varphi>0\\
\sqrt{A_{N}}\exp\Big\{+\i \frac{N}{2}\varphi -\sum_{n=1}^{\infty
}\frac{t_{n}^{*}\mathrm{e}^{-\i n\varphi}}{n}\Big\} & \quad\mathrm{Im}
\,\varphi<0
\end{array}
\right. \,.
\end{equation}
That representation of zeta by an exponentiated infinite Fourier sum
allows us to write the reciprocal $1/\zeta$ as
\begin{equation}
\label{invzeta}
\frac{1}{\zeta(\varphi)}=\left\{
\begin{array}
[c]{ll}
\sqrt{A_{N}}\exp\Big\{+\i \frac{N}{2}\varphi +\sum_{n=1}^{\infty
}\frac{t_{n}\mathrm{e}^{\i n\varphi}} {n}\Big\} & \quad\mathrm{Im}
\,\varphi>0\\
\sqrt{A_{N}^{*}}\exp\Big\{-\i \frac{N}{2}\varphi +\sum
_{n=1}^{\infty}\frac{t_{n}^{*}\mathrm{e}^{-\i n\varphi}} {n}\Big\} &
\quad\mathrm{Im}\,\varphi<0
\end{array}
\right. \,.
\end{equation}
The innocent looking sign changes relative to (\ref{altzeta}) have in fact
drastic consequences. Clearly, the Fourier series expansion of $1/\zeta$
cannot be a finite-order polynomial in $\mathrm{e}^{\pm\i\varphi}$ but rather
is an infinite series,
\begin{equation}
\label{invzetadown}
\frac{1}{\zeta(\varphi)}=\left\{
\begin{array}
[c]{ll} \sqrt{A_{N}}\,\mathrm{e}^{\i N\varphi/2}\;
 \sum_{n=0}^{\infty}\tilde
{A}_{n}\,\mathrm{e}^{\i n\varphi} & \quad\mathrm{Im}\,\varphi>0\\
\sqrt{A_{N}^{*}}\,\mathrm{e}^{-\i N\varphi/2 }
\;\sum_{n=0}^{\infty }\tilde{A}_{n}^{*}\,\mathrm{e}^{-\i n\varphi}
& \quad\mathrm{Im}\,\varphi<0
\end{array}
\right. \,,
\end{equation}
wherein the Fourier coefficients $\tilde{A}_{n}$ differ from the secular
coefficients $A_{n}$ only by the absence of the sign factor in (\ref{An}),
\begin{equation}
\label{tildeAn}\tilde{A}_{n}=\sum_{\vec{v}}^{\{n=L(\vec{v})\}} \prod_{l\geq
1}\frac{t_{l}^{v_{l}}}{l^{v_{l}}v_{l}!}\,.
\end{equation}

\section{Generating function}

\label{generatingfunction}

In analogy to standard practice for energy spectra
we introduce the
multiplicative combination of four spectral determinants,
\begin{equation}
\label{Z}Z=\left\langle \frac{\zeta(\varphi_{C})\zeta(\varphi_{D})}
{\zeta(\varphi_{A})\zeta(\varphi_{B})} \right\rangle \,.
\end{equation}
The angular brackets denote two averages. One is over $2\pi$-interval of a
real center phase $\phi$ defined through
\begin{eqnarray}
\label{eq:13}\varphi_{A/C}=\phi+\frac{e_{A/C}}{N}\,, \qquad\varphi_{B/D}%
=\phi-\frac{e_{B/D}}{N}\,.
\end{eqnarray}
A second average is, in principle, necessary over a small offset
window \cite{Haake10}. The generating function $Z$ deserves
interest since it yields the two-point function of the level
density through two derivatives, see \cite{Haake10} and
\ref{2deriv}.

To make the average
$\langle\cdot\rangle=(2\pi)^{-1}\int_{0}^{2\pi}d\phi (\cdot)$ well
defined, we must assign nonvanishing imaginary parts of opposite sign
to the phase arguments in the denominator on the r.~h.~s.~of
(\ref{Z}).  Therefore, we endow the offset variables $e_{A/B}$ with
positive imaginary parts. On the other hand, the phase arguments of
the numerator zetas are unrestricted by the definition of $Z$.

To prepare for the semiclassical analysis of the generating function
$Z$ we represent the two inverse zetas in the definition (\ref{Z}) by
the infinite series (\ref{invzetadown}) but employ the Riemann-Siegel lookalike
(\ref{zetaeven}) in the numerator. The generating function thus
becomes a sum of nine terms,
\begin{eqnarray}\fl
\label{Zij}
Z =\sum_{i,j=+,-,0}Z_{ij}\,, \qquad
Z_{ij}=\big\langle\zeta(\varphi_{A})^{-1}\zeta(\varphi_{B})^{-1}
\zeta_i(\varphi_{C})\zeta_j(\varphi_{D})\big\rangle\,.
\end{eqnarray}
The symmetries
\begin{eqnarray}
\label{Zsymmetries}
Z_{+-}(e_{A},e_{B},e_{C},e_{D})&=Z_{-+}(e_{A},e_{B},-e_{D},-e_{C})\\
Z_{++}(e_{A},e_{B},e_{C},e_{D})&=Z_{--}(e_{A},e_{B},-e_{D},-e_{C})\nonumber\\
Z_{+0}(e_{A},e_{B},e_{C},e_{D})&=Z_{0+}(e_{A},e_{B},-e_{D},-e_{C})\nonumber\\
Z_{0-}(e_{A},e_{B},e_{C},e_{D})&=Z_{-0}(e_{A},e_{B},-e_{D},-e_{C})\nonumber
\end{eqnarray}
show that only five components, say,
$Z_{-+},Z_{--},Z_{-0},Z_{0+},Z_{00}$ are independent. Explicit
expression for the $Z_{ij}$ will be written as need arises. For now,
we just note the example
\begin{eqnarray}
Z_{-+}  & =\mathrm{e}^{\frac{\i}{2}(e_{A}+e_{B}-e_{C}-e_{D})}\sum_{a,b=0}^\infty
\sum_{c,d=0}^{\frac{N}{2}-1}\tilde{A}_{a}\tilde{A}_{b}^{\ast}A_{c}A_{d}
^{\ast}\label{Z-+}\\
& \qquad\times\langle\mathrm{e}^{\i(a-b+c-d)\phi}\rangle\,\mathrm{e}
^{\frac{\i}{N}(ae_{A}+be_{B}+ce_{C}+de_{D})}\,;\nonumber
\end{eqnarray}
herein, the center-phase average yields the restriction $a+c=b+d$ such
that only the offset phases remain as variables. The summations over
the integers $c,d$ go from $0$ to $\frac{N}{2}-1$, due to the
definitions of $\zeta_\pm$. Clearly then, $Z_{-+}$ is a polynomial of
order $\frac{N}{2}$ both with respect to $\e^{-\i e_C/N}$ and $\e^{-\i
  e_D/N}$, without zero-order terms in either variable. It is now
convenient to introduce projectors $\mathcal{P}_-^{C}$ and
$\mathcal{P}_-^{D}$ leaving the powers $\mathrm{e}^{\i \nu e_{C}/N}$
and $\mathrm{e}^{\i \nu e_{D}/N}$ with $-\frac{N}{2}\le\nu<0$
unchanged and killing all others. Under the protection of the product
$\mathcal{P}_-^C\mathcal{P}_-^D$ the sums over the integers $c,d$ can
be extended to go from $0$ to infinity, like the sums over $a,b$. With
the help of the expansion coefficients (\ref{An},\ref{tildeAn}) we get
$Z_{-+}$ represented by
\begin{eqnarray}
\fl
Z_{-+}=\mathcal{P}_-^C\mathcal{P}_-^D\Bigg(\mathrm{e}^{\frac{\i}{2}(e_{A}+e_{B}-e_{C}-e_{D})}
\sum_{\vec{a},\vec{b},\vec{c},\vec{d}}^{L(\vec{a})+L(\vec{c})
=L(\vec{b})+L(\vec{d})}\label{Z-+Fourier}\\\hspace{-2em}
 \prod_{n=1}^{\infty}\Big[\Big(\frac{t_{n}}{n}\Big)^{a_{n}+c_{n}}%
\Big(\frac{t_{n}^{\ast}}{n}\Big)^{b_{n}+d_{n}}\frac{(-1)^{c_{n}+d_{n}}}%
{a_{n}!b_{n}!c_{n}!d_{n}!}\,\mathrm{e}^{\i\frac{n}{N}(a_{n}e_{A}%
+c_{n}e_{C}+b_{n}e_{B}+d_{n}e_{D})}\Big]\Bigg)\,.\nonumber
\end{eqnarray}

Similarly, we define projectors
$\mathcal{P}_0^C,\,\mathcal{P}_0^D$ which leave unchanged the
phase independent term in the respective zetas while killing all
$\mathrm{e}^{\i \nu e_{C}/N}$ and $\mathrm{e}^{\i \nu e_{D}/N}$
with $\nu\neq 0$. We then have
$Z_{00}=\mathcal{P}_0^C\mathcal{P}_0^D\big(\ldots\big)$, \
$Z_{0+}=\mathcal{P}_0^C\mathcal{P}_-^D\big(\ldots\big)$, and
$Z_{-0}=\mathcal{P}_-^C\mathcal{P}_0^D\big(\ldots\big)$ with the
parentheses the same as for $Z_{-+}$ above.  The analogous series
for $Z_{--}$ reads\newpage
\begin{eqnarray}\fl
Z_{--}=\mathcal{P}_-^C\mathcal{P}_+^D\Bigg\{\mathrm{e}^{\frac{\i}{2}(e_{A}+e_{B}-e_{C}+e_{D})}
A_{N}^{\ast}\sum_{\vec{a},\vec{b},\vec{c},\vec{d}}^{L(\vec
{a})+L(\vec{c})+L(\vec{d})=L(\vec{b})+N}\label{Z--Fourier}\\\hspace{-2em}
\prod_{n=1}^{\infty}\Big[\Big(\frac{t_{n}}{n}\Big)^{a_{n}+c_{n}+d_{n}
}\Big(\frac{t_{n}^{\ast}}{n}\Big)^{b_{n}}\frac{(-1)^{c_{n}+d_{n}}}{a_{n}
!b_{n}!c_{n}!d_{n}!}\,\mathrm{e}^{\i\frac{n}{N}(a_{n}e_{A}+c_{n}
e_{C}+b_{n}e_{B}-d_{n}e_{D})}\Big]\Bigg\}\,\nonumber
\end{eqnarray}
where the projector $\mathcal{P}_+^D$ leaves unchanged the powers
$\mathrm{e}^{+\i \nu e_{D}/N}$ with $0<\nu\leq\frac{N}{2}$ and kills
all others.

There is a price to pay for the introduction of the infinite sums over
the vectors $\vec{c},\vec{d}$ in
(\ref{Z-+Fourier},\ref{Z--Fourier}). To ensure convergence we must
allow for infinitesimal positive imaginary parts of $e_{C/D}$, as
already done for $e_{A/B}$.

\section{\bigskip Semiclassics for maps}\label{semiclassics driven}

\subsection{Trace formula}

We now foray into asymptotics by invoking the Gutzwiller trace
formula. For the traces $t_{n}$ of Floquet maps of periodically
driven systems we have \cite{Haake10}
\begin{equation}
t_{n}\sim\sum\frac{n}{\sqrt{|\det(M-1)|}}\,\mathrm{e}^{\i S/\hbar
}\,,\label{maptrace}%
\end{equation}
a sum over period-$n$ orbits. The contribution of each orbit involves
a phase factor where the action $S$ (defined to include the Maslov
phase and measured in units of Planck's constant) appears as the
phase. That $S$  is the 'time dependent' action which generates
the periodic orbit in $n$ steps of the classical map. The prefactor involves the primitive period $n$ and accounts
for the instability of the orbit in terms of the so called monodromy
matrix $M$. For the sake of simplicity, we shall treat periodically
driven single-freedom systems. We thus have a single positive
Lyapounov rate $\lambda$ to reckon with for each periodic orbit. For
beauty, we refrain from putting subscripts distinguishing different
period-$n$ orbits on the action and the monodromy matrix.

The number $\mathcal{N}(n)$ of period-$n$ orbits is known to
asymptotically grow exponentially with the period $n$
\cite{Margu04,Bowen72,Hanna84}. On average over a small window
$\Delta n$ of periods centered at $\overline{n}$, such that
$1\ll\Delta n\ll\overline{n}$, one has
\begin{equation}\label{HOdA}
\langle \mathcal{N}(n)\rangle=\mathrm{e}^{n\lambda\tau}/n\,.
\end{equation}
The finiteness of the number of period-$n$ orbits makes the trace
formula (\ref{maptrace}) a well defined object not in need of any
regularization, quite in contrast to the more widely known one for
the density of energy levels of autonomous dynamics (see next
section).

Long orbits dominate the semiclassical asymptotics, and therefore we have
written the trace formula (\ref{maptrace}) without regard for $r$-fold
traversals of orbits with primitive period $n_{0}=n/r$. Assuming ergodicity we
encounter one and the same Lyapounov rate for all long periodic orbits. The
eigenvalues $\mathrm{e}^{\pm n\lambda\tau}$ of the monodromy matrix give the
stability prefactor $|\det(M-1)|^{-1/2}\sim\mathrm{e}^{-n\lambda\tau/2}$ such
that the powers of the traces $t_{n},t_{n}^{\ast}$ in the Fourier series
(\ref{Z-+Fourier}) and (\ref{Z--Fourier}) for $Z_{-+}$ and $Z_{--}$ take the form
\begin{eqnarray}\label{GutzipowersZ-+}
\fl
&\Big(\frac{t_{n}}{n}\!\Big)^{\!a_{n}+c_{n}}\!\Big(\frac{t_{n}^{\ast}}{n}\!\Big)^{\!b_{n}+d_{n}}
 =\mathrm{e}^{-\frac{n\lambda\tau}{2}(a_{n}+b_{n}+c_{n}+d_{n})}
\Big(\!\textstyle{\sum}\mathrm{e}^{\frac{\i}{\hbar}S}\!\Big)^{\!a_{n}+c_{n}}\!
\Big(\!\sum\mathrm{e}^{-\frac{\i}{\hbar}S}\!\Big)^{\!b_{n}+d_{n}},\\
\fl
&\Big(\frac{t_{n}}{n}\!\Big)^{\!a_{n}+c_{n}+d_n}\!\Big(\frac{t_{n}^{\ast}}{n}\!\Big)^{\!b_{n}}
 =\mathrm{e}^{-\frac{n\lambda\tau}{2}(a_{n}+b_{n}+c_{n}+d_{n})}
\Big(\!\textstyle{\sum}\mathrm{e}^{\frac{\i}{\hbar}S}\!\Big)^{\!a_{n}+c_{n}+d_n}\!
\Big(\!\sum\mathrm{e}^{-\frac{\i}{\hbar}S}\!\Big)^{\!b_{n}}\label{GutzipowersZ--}.
\end{eqnarray}

\subsection{Diagonal approximation, unitary class}

\label{diagapproxautonomous}

We now focus on the diagonal approximation for the unitary
symmetry class: only those terms from the multinomial expansion of
the powers of periodic-orbit sums in
(\ref{GutzipowersZ-+},\ref{GutzipowersZ--}) are kept in which each
phase factor $\mathrm{e}^{\i S/\hbar}$ pairs up with its complex
conjugate such that all dependence on the actions cancels. We thus
immediately get $\left[\left(t_n\right)^m
\left(t_n^*\right)^{m'}\right]_{\rm diag}\propto \delta_{mm'}$.
For fixed $m$, any choice of $m$ action-phase factors
$\mathrm{e}^{\i S/\hbar}$ comes with $m!$ possibilities to pick
the matching inverses. With any of these choices the $m$-fold
summation over periodic orbits leads to the $m$-th power of the
diagonal approximation for the product $t_n t_n^*$ of just two
traces,
\begin{equation}
  \label{diagGauss}
\fl\left[t_n^m \left(t_n^*\right)^{m'}\right]_{\rm
diag}=\delta_{mm'}\,m!\,\left[t_nt_n^*\right]_{\rm diag}^m
=\delta_{mm'}m!\,\left(n^2\e^{-n\lambda\tau}\mathcal{N}(n)\right)^m\,,
\end{equation}
a behavior reminiscent of Gaussian statistics.

Using the result (\ref{diagGauss}) for $Z_{-+}^{\rm diag}$ we face
\begin{eqnarray}
&
Z_{-+}^{\mathrm{diag}}=\mathcal{P}_-^C\mathcal{P}_-^D\big(\ldots\big)
\end{eqnarray}
with the building block
\begin{eqnarray}\nonumber
\fl
\Big(\ldots\Big)=\mathrm{e}^{\frac{\i}{2}(e_{A}+e_{B}-e_{C}-e_{D})}
\prod_{n=1}^{\infty}\sum_{m=0}^\infty\sum
_{a_{n},b_{n},c_{n},d_{n}}^{a_{n}+c_{n}=b_{n}+d_{n}=m}\,\frac{(-1)^{c_{n}
+d_{n}}}{m!}\left(\e^{-n\lambda\tau}\mathcal{N}(n)\right)^m\\
 \qquad\qquad\times\,\,\left({m \atop a_n}\right)\left({m\atop b_n}\right)
\e^{\i\frac{n}{N}(a_{n}e_{A}+c_{n}e_{C}+b_{n}e_{B}+d_{n}e_{D})}\,.
\end{eqnarray}
We here do two binomial sums and sum over $m$,
\begin{eqnarray}
\fl\nonumber \Big(\ldots\Big)
=\mathrm{e}^{\frac{\i}{2}(e_{A}+e_{B}-e_{C}-e_{D})}
\exp\sum_{n=1}^{\infty}\e^{-n\lambda\tau}\mathcal{N}(n)
\big(\mathrm{e}^{\i\frac{n}{N}e_{A}}-\mathrm{e}^{\i\frac{n}{N}e_{C}}\big)
\big
(\mathrm{e}^{\i\frac{n}{N}e_{B}}-\mathrm{e}^{\i\frac{n}{N}e_{D}}
\big)\,.
\end{eqnarray}

In the summand of the sum over the orbit periods $n$ within the
foregoing expression, the factor $\mathcal{N}(n)$ fluctuates about the
average given by the HOdA sum rule (\ref{HOdA}). Inasmuch as the
cofactor depending on the offset variables $e_{A/B/C/D}$ varies
smoothly with $n$ on the scale $\Delta n$ of validity of the HOdA sum
rule, we can now replace the fluctuating $\mathcal{N}(n)$ by its
average, $\e^{-n\lambda\tau}\mathcal{N}(n)\to 1/n$. By this 'selfaveraging' in the
semiclassical limit, the generating function becomes smooth without any
average beyond the one over the center phase $\phi$.

Finally, using $\sum_{1}^{\infty}\frac{\mathrm{e}^{\i nx}}{n}
=-\ln(1-\mathrm{e}^{\i x})$ we get\footnote{Strictly speaking, we are
  not allowed to use the exponential proliferation factor for the
  average number of period-$n$ orbits for small periods $n$. However,
  the error thus committed is of order $(n_{0}/N)^{2}$ with $n_{0}$
  some period above which the HOdA sum rule is reliable. For a rough
  argument, assume the offsets of order unity and consider
  $\sum_{n=1}^{n_{0}
  }\big(\mathrm{e}^{\i\frac{n}{N}e_{A}}-\mathrm{e}^{\i\frac{n}{N}e_{C}
  }\big)\big (\mathrm{e}^{\i\frac{n}{N}e_{B}}-\mathrm{e}^{\i\frac{n}
    {N}e_{D}}\big)/n\sim(e_{A}-e_{C})(e_{B}-e_{D})\sum_{n=1}^{n_{0}}\frac{n}
  {N^{2}}\propto(\frac{n_{0}}{N})^{2}$; the argument is easily
  modified to allow for the period $2\pi N$ of the offset variables.}
 \begin{equation}\label{(...)}
 \Big(\ldots\Big)  =\e^{\frac{\i}{2}(e_{A}+e_{B}-e_{C}-e_{D})}\;
\frac{1-\e^{\frac{\i}{N}(e_A+e_D)}}{1-\e^{\frac{\i}{N}(e_A+e_B)}}
\frac{1-\e^{\frac{\i}{N}(e_B+e_C)}}{1-\e^{\frac{\i}{N}(e_C+e_D)}}\,.
\end{equation}

Let us first throw  a glance at $Z_{--}$ as given by
(\ref{Z--Fourier},\ref{GutzipowersZ--},\ref{diagGauss}). The phase
matching required by the diagonal approximation yields the
restriction $a_{n}+c_{n}+d_{n}=b_{n}$ which contradicts the one
obtained from the center-phase average,
$\sum_{n}n(a_{n}+c_{n}+d_{n}-b_{n})=N $. We conclude that
$Z_{--}^{\rm diag}$ vanishes. Due to the symmetry
(\ref{Zsymmetries}) we have $Z_{++}^{\rm diag}=0$ as well. The
full generating function thus comes out as
\begin{eqnarray}
 \fl Z^{\rm diag}=Z_{00}^{\rm diag}
  +\big(Z_{-+}^{\rm diag}+Z_{0+}^{\rm diag}+Z_{-0}^{\rm diag}\big)
  +\big(Z_{+-}^{\rm diag}+Z_{+0}^{\rm diag}+Z_{0-}^{\rm diag}\big)\,.
\end{eqnarray}
The first four terms are obtained from $\left(\ldots\right)$ by
applying respectively the projectors
$\mathcal{P}_0^C\mathcal{P}_0^D,\mathcal{P}_-^C\mathcal{P}_-^D,\mathcal{P}_0^C\mathcal{P}_-^D,\,
\mathcal{P}_-^C\mathcal{P}_0^D$; the remaining ones follow from the
symmetry conditions (\ref{Zsymmetries}). To let these projectors do
their job we expand as $(1-\e^{\frac{\i}{N}
  (e_{C}+e_{D})})^{-1}=\sum_{\nu=0}^\infty \e^{\frac{\i\nu}{N}
  (e_{C}+e_{D})}$. After elementary calculations given in \ref{projector} we
obtain a polynomial of order $\frac{N}{2}$ in $e^{\pm
  \frac{\i}{N}e_C}$ and $e^{\pm\frac{\i}{N}e_D}$. Amazingly, the final
expression coincides with the sum of $\left(\ldots\right)$ and its
Riemann-Siegel complement (obtained by the replacement $e_C,e_D\to
-e_D,-e_C$),
\begin{eqnarray}
  \label{Zirni}
  Z^{\rm diag}&=&\e^{\frac{\i}{2}(e_{A}+e_{B}-e_{C}-e_{D})}\;
\frac{1-\e^{\frac{\i}{N}(e_A+e_D)}}{1-\e^{\frac{\i}{N}(e_A+e_B)}}
\frac{1-\e^{\frac{\i}{N}(e_B+e_C)}}{1-\e^{\frac{\i}{N}(e_C+e_D)}}\\
&&+\,\e^{\frac{\i}{2}(e_{A}+e_{B}+e_{C}+e_{D})}\;
\frac{1-\e^{\frac{\i}{N}(e_A-e_C)}}{1-\e^{\frac{\i}{N}(e_A+e_B)}}
\frac{1-\e^{\frac{\i}{N}(e_B-e_D)}}{1-\e^{-\frac{\i}{N}(e_C+e_D)}}\nonumber\,.
\end{eqnarray}
Indeed, the denominator $(1-\e^{\frac{\i}{N} (e_{C}+e_{D})})$ cancels
from the sum of the two terms in (\ref{Zirni}) such that the
polynomial character mentioned arises.

Most remarkably, we have arrived at the exact generating function for
the circular unitary ensemble \cite{Conre07} for
\textit{finite} $N$. In particular, the periodicity in the offset
phases with period $\propto N$ is a nice finite-size effect for
Floquet operators of periodically driven dynamics.

Of course, in the limit $N\to \infty$ we get the generating function
describing correlation decay at finite offsets
\begin{eqnarray}
\lim_{N\rightarrow\infty}Z^{\mathrm{diag}} \,  = \;\;&
\mathrm{e}^{\i
(e_{A}+e_{B}-e_{C}-e_{D})/2}\,\frac{(e_{A}+e_{D})(e_{B}+e_{C})}{(e_{A}%
+e_{B})(e_{C}+e_{D})}\label{eq:7a}\\
& -\mathrm{e}^{\i(e_{A}+e_{B}+e_{C}+e_{D})/2}\,\frac{(e_{A}-e_{C}%
)(e_{B}-e_{D})}{(e_{A}+e_{B})(e_{C}+e_{D})}\nonumber\,,
\end{eqnarray}
as for the Gaussian unitary ensemble.

The case of odd dimension $N$ leads to the same result. The pertinent
calculation differs in technical details not worth being spelled out
here.

\subsection{Off-diagonal corrections}
\label{sec:offdiag}

Inasmuch as the foregoing diagonal approximation already gives the
exact result for the unitary symmetry class, all off-diagonal
contributions must cancel. There is nothing much to add to the
 previous work for flows of autonomous dynamics as far as
the limit $N\to\infty$ is concerned.

Some nontrivial modifications are needed for finite $N$
and the periodicity of $Z$. While a full proof of the cancelation
of off-diagonal corrections will be reserved for a separate paper,
we would like to at least give some hints here.

As previously, quadruples of non-ordered sets of periodic orbits
(pseudo-orbits) come into play such that the pseudo-orbits within
each quadruple have action differences of the order of Planck's
constant (see the next section for some more details). The
(Feynman type) diagrams characterizing those quadruples remain
unchanged, and so is the ordering of the diagrams in terms of the
number of vertices ($V$) and links ($L$) by increasing values of
$L-V$. The rules for associating analytic expressions with
diagrams are modified. Straightforward computer-based summations
yield order-by-order cancelation of the corrections with
$L-V=1,2,\ldots,6$ which altogether involve 7404 non-equivalent
diagrams.

\section{Stroboscopic maps for autonomous flows}\label{mapsforflows}

The rigorous Riemann-Siegel lookalike for finite dimensional
unitary matrices can also be made available for autonomous flows,
by choosing a suitable stroboscopic description. We propose to
show that the status of previous work benefits from that
modification. In particular, the diagonal approximation for the
unitary symmetry class will again lead us to the exact finite-$N$
result (\ref{Zirni}), without needing large imaginary parts for the
offset variables $e_{C/D}$. The periodicity in the
offset variables is in this case an artefact of the stroboscopic
description; only the limit $N\to\infty$ is of interest here.

\subsection{Classical strobe period}
\label{classicalperiod}

We consider a small stretch of the spectrum of a time independent
Hamiltonian containing $N\gg 1$ levels $\{E_{1}\leq
E_{2}\leq\ldots\leq E_{N}\}$ and the pertinent time evolution operator
over a certain strobe period $\tau$, to be called Floquet operator,
$F=\sum_{\mu=1}^N {\mathrm e}^{-\i
  E_\mu\tau/\hbar}|\mu\rangle\langle|\mu|$. Choosing the strobe period
such that the eigenphases $\varphi_{\mu}=E_{\mu}\tau/\hbar$ of $F$
just fill the interval $[0,2\pi]$ once, namely
$\tau=2\pi\hbar/(E_{N}-E_{1})$, the increasing order of the
eigenenergies carries over to the eigenphases. For large $N$, the
spectral fluctuations of the Hermitian Hamiltonian and the unitary
Floquet operator become identical, since the neighborship of
$\varphi_1$ and $\varphi_N$ loses any significance.

Three time scales are of importance for us, the Heisenberg time $T_{H}=N\tau$,
the Ehrenfest time $T_{E}=\lambda^{-1}\ln(S/\hbar)$ with $\lambda$ the
Lyapounov rate and $S$ some classical action scale, and the strobe period
$\tau$. These times are ordered as $\tau\ll T_{E}\ll T_{H}$. In fact, we may
take the strobe period to be a classical time, independent of Planck's
constant; in view of $T_{H}\sim\hbar^{-f+1}$, classical $\tau$ means
$N\sim\hbar^{-f+1} $. On the other hand, classical $\tau$ means an overall
energy span $E_{N}-E_{1}=\mathcal{\ O}(\hbar)$, and therefore the mean level
density therein, $\overline\rho=N/(E_{N}-E_{1})$, needs no unfolding.

\subsection{Trace formula}
\label{traceformula}

We adopt the Gutzwiller formula for the density of energy levels
\cite{Haake10,Gutzw90},
\begin{equation}
  \label{Gutzidensity}
  \rho(E)\sim
\overline\rho+\frac{1}{2\pi\hbar}\sum\frac{T}
{\sqrt{\det|M-1|}}(\e^{\i S^0/\hbar}+\e^{-\i S^0/\hbar})\,.
\end{equation}
Here the sum is over periodic orbits on the energy shell $E$ and
$S^{0}$ is the energy dependent action (sometimes called reduced
action) which as a generating function generates periodic orbits
for fixed energy $E$; finally, $T$ denotes the primitive period. A
Legendre transformation connects the energy dependent action and
its time dependent counterpart as $S=S^{0}-ET$. We again employ
$\det|M-1|\sim \e^{\lambda T}$. The traces of the $N\times N$
Floquet matrix $t_{n}=\sum_{k=1}^N e^{-in\tau E_k/\hbar}$ can be
written
\begin{eqnarray}\label{window}
\fl t_n=\int dE'\left[\theta(E'-E+\Delta E/2)-\theta(E'-E-\Delta
E/2)\right] \rho(E')\e^{-\i E'n\tau/\hbar}.
\end{eqnarray}
The step functions select the energy interval $[E-\Delta E/2,E+\Delta E/2]$
filled by the spectrum,  such that $\tau\Delta E/\hbar=2\pi$.
Semiclassically for $n>0$ we have,
\begin{equation}
\fl t_{n}\sim\frac{1}{2\pi\hbar}\sum\int_{E-\Delta E/2}^{E+\Delta
E/2}dE'T \e^{-\i E'n\tau/\hbar -\lambda T/2} \big(\e^{\i
S^{0}(E^{\prime})/\hbar}+ \e^{-\i S^{0}(E^{\prime})/\hbar}\big).
\label{eq:26}
\end{equation}
The reference energy at the mid-point of the spectrum
will henceforth be set to zero, $E=0$. To check how strongly the
phase of the exponentials within the integrand varies across that
interval, we expand
$[S^0(E^{\prime})-E^{\prime}n\tau]/\hbar=\big[S^{0}(0)
+(T-n\tau)E^{\prime}+\frac{1}{2}S^{\prime\prime}(0)(E^{\prime
})^{2}+\ldots\big]/\hbar$. The third term in that expansion
defines a width $\propto\sqrt{\hbar}$ larger than the span of the
spectrum such that the term can be dropped. The second term of the
expansion, on the other hand, varies by $\Delta\phi\equiv2\pi(T%
-n\tau)/\tau$ across the spectral range. We can conclude that the
term $\mathrm{e}^{-\i S^{0}/\hbar}$ is negligible for $n>0$ while
$\mathrm{e}^{\i S^{0}/\hbar}$ plays no role for $n<0$. Inasmuch as
we need the traces $t_{n} $ only for positive $n$ we have,
\begin{eqnarray}
t_{n}  & \sim\sum\frac{T}{\tau}
\e^{\i S^0/\hbar-\lambda T/2}\frac{1}{\Delta E}\int_{-\Delta E/2}^{\Delta E/2}dE'
\e^{\i(T-n\tau) E'/\hbar}\nonumber\\
& \sim\sum\frac{T}{\tau}\e^{\i S^{0}/\hbar-\lambda T/2}
\mathrm{sinc}\left( \frac{T}{\tau}-n\right) \,.\label{flowtrace}
\end{eqnarray}
The function $\mathrm{sinc}\,x=\sin\pi x/\pi x$ favors orbits whose
periods $T$ do not differ from $n\tau$ by more than a few $\tau$
(see next subsection). The sharpness of the energy and the
uncertainty of the orbit periods of the order $\tau$ are the
important differences of the present trace formula to the one
valid for periodically driven systems, see Eq.~(\ref{maptrace}).
As a consequence of that difference it will turn out necessary to
introduce a certain regularization below.

\subsection{Diagonal approximation, unitary symmetry}
\label{diagapproxdriven} We begin with calculating the product of
$ t_{n}t_{n}^{\ast } $ in the diagonal approximation following
from the periodic-orbit expansion (\ref{flowtrace}). Denoting by
$T_{0}$ the smallest period above which the average number  of
periodic orbits with periods in $[T,T+dT$ is given by the
exponential proliferation law $ dTe^{\lambda T}/T$ we have,
\begin{eqnarray}\label{intfortraces}
\left[ t_{n}t_{n}^{\ast }\right] _{\mathrm{diag}} &=&\sum\left(
\frac{T}{\tau }\right) ^{2} \e^{-\lambda T}\mathrm{sinc}^2 \left(
\frac{T}{\tau }-n\right)
\nonumber \\
&\sim &\int_{T_{0}}^{\infty }\frac{dT}{T}\left( \frac{T}{\tau }\right) ^{2}%
\mathrm{sinc}^2\left( \frac{T}{\tau }-n\right)\,.
\end{eqnarray}%
This expression formally diverges at the upper limit due to slow decay
of the sinc function at large values of the argument. It is here that
the regularization announced just above becomes necessary. To that
end, suppose that in (\ref{window}) we smoothen the cut-offs at
$E'=\pm \Delta E/2$ such that transitions from 0 to 1 and back take
place through intervals of the order of the mean level spacing
$\delta E=\Delta E/N$. That smoothing cannot noticeably change traces of the
Floquet matrix. However, in the periodic-orbit expansion the
contributions with periods significantly larger than the Heisenberg
time will be suppressed\footnote{We can, e.g., replace the thetas in
  (\ref{window}) by the error functions
  $\frac{1}{2}\left[1+\mathrm{erf}\left(\pi\frac{E'-E\pm\frac{\Delta
          E}{2}}{\delta E}\right)\right]$. The periodic-orbit
  expansion of traces (\ref{eq:26}) will then acquire the Gaussian
  factor $\e^{-T^2/T_H^2},\quad T_H=2\pi\hbar/\delta E$, and the
  result (\ref{trace_regularized}) will be correct to within
  corrections of order of $\ln N$ which may be neglected compared with
  relevant values of $n$.}.
The integral (\ref{intfortraces}) then becomes finite with the main
contribution from the region where the argument of sinc is close to
zero such that $T$ can be replaced by $ n\tau$. After that the
integration interval can be safely extended to $\left[ -\infty ,\infty
\right] $ with the result
\begin{equation}\label{trace_regularized}
\left[ t_{n}t_{n}^{\ast }\right] _{\mathrm{diag}}\sim n\int_{-\infty
}^{\infty }
\frac{dT}{\tau }\mathrm{sinc}^2\left( \frac{T}{\tau }%
-n\right)  =n.
\end{equation}

The diagonal approximation for products of higher powers of traces
$t_n^m(t_n^*)^k$ can be expressed in terms of $\left[ t_{n}t_{n}^{\ast
  }\right] _{\mathrm{diag}}$ in the same way as for periodically
driven systems, and we recover Eq.  (\ref{diagGauss}).

Likewise, all subsequent steps of the diagonal approximation in
Sect.~\ref{diagapproxautonomous} are unchanged such that the
generating function (\ref{Zirni}) is rederived. In the present case of
autonomous flows, only the limit of small offsets, $e_{A/B/C/D}/N \to
0$ yielding (\ref{eq:7a}), is of physical interest. Two significant
advantages over the derivation in \cite{Heusl07,Muell09, Haake10} are
worth noting.  First, we now need only infinitely small imaginary parts
of the offset phases $e_{C/D}$.  Second, as long as $N$ is finite,
the finiteness of the Fourier series in the variables $e_{C/D}$ is
rigorously preserved; while the Riemann-Siegel components of $Z$
individually have non-vanishing contributions from pseudo-orbits with
periods $T>T_H/2$, their sum enjoys cancellation of these
contributions.

\subsection{Quadruples of pseudo-orbits for $Z$}

We recall the Riemann-Siegel lookalike (\ref{zetaeven}) and the
ensuing decomposition (\ref{Zij}) of the generating function. It
is now convenient to write the component $Z_{-+}$ in the form
\begin{eqnarray}
  \label{Z1flow}
  Z_{-+}&=&\Big\langle \mathcal{P}_-^C\mathcal{P}_-^D\,
  \e^{\frac{\i}{2}N(\varphi_A-\varphi_C+\varphi_D-\varphi_B)}\\
  \nonumber
        &&\quad \times\exp\sum_{n=1}^\infty\Big\{
          \frac{t_n}{n}\big(\e^{\i n\varphi_A}-\e^{\i  n\varphi_C}\big)+
          \frac{t_n^*}{n}\big(\e^{-\i n\varphi_B}-\e^{\i
            n\varphi_D}\big)\Big\}
     \Big\rangle
\end{eqnarray}
which is easily seen to be correct when using the projectors
$\mathcal{P}_-^C,\mathcal{P}_-^D$ in (\ref{altzeta}).

After invoking the trace formula (\ref{flowtrace}) the sums over
the traces in (\ref{Z1flow}) are expressed in terms of the
periodic orbits,
\begin{equation}
  \label{Fouriertnovern}
\sum_{n=1}^\infty\frac{t_n}{n}\e^{\i n\varphi }\sim\sum\e^{\i S^0/\hbar-\lambda T/2}
\sum_{n=1}^\infty\frac{T}{n\tau}\e^{\i n\varphi}
 \mathrm{sinc}\left( \frac{T}{\tau}-n\right)\,.
\end{equation}
Inasmuch as orbit periods of the order of the Heisenberg time $N\tau$
dominate we may replace the factor $\frac{T}{n\tau}$ with
unity, accepting a relative error of the order $\frac{1}{N}$. The
lower limit of the sum over $n$ can then be shifted to
$-\infty$.

Next, we employ the discrete Fourier transform
\begin{equation}
  \label{Fouriersinc}
\sum_{n=-\infty}^{\infty}\e^{\i n\varphi}\mathrm{sinc}(x-n)=\e^{\i
x\{\varphi\}}
\end{equation}
where $\{\varphi\}$ is the $(2\pi)$-periodic sawtooth function
equalling $\varphi$ in the interval $[-\pi, \pi]$. For use further
below we note right away that restricting the sum over $n$ to the
range $|n|\leq \frac{N}{2}$ means putting a soft cutoff on the
parameter $x$ to $|x|<\frac{N}{2}$, because for fixed $n$ and varying
$x$ the function ${\rm sinc}(x-n)$ is weakly localized near $x=n$.  At
any rate, the sum in (\ref{Fouriertnovern}) can be approximated as
\begin{equation}
  \label{Fouriertnovern2}
   \sum_{n=1}^\infty\frac{t_n}{n}\e^{\i n\varphi }\sim\sum
\e^{\i  S^0/\hbar-\lambda T/2+\i \{\varphi\}T/\tau}
\end{equation}
and herein the truncation $n\leq \frac{N}{2}$ would entail the soft
cutoff $\frac{T}{\tau}<\frac{N}{2}$.

Four such sums appear exponentiated in the expression (\ref{Z1flow})
for $Z_1$. We Taylor expand all four pertinent exponentials and so
obtain a fourfold sum over pseudo-orbits. Neglecting orbit repetitions
and again representing the four phases $\varphi_{A/B/C/D}$ as in
(\ref{eq:13}) we find \cite{Heusl07,Muell09,Haake10}%

\begin{eqnarray}
 \label{Z1flow2}\hspace{-5em}
 Z_{-+}\sim \mathcal{P}_-^C\mathcal{P}_-^D\,\e^{\i
(e_A+e_B-e_C-e_D)/2}
\sum_{\mathcal{A},\mathcal{B},\mathcal{C},\mathcal{D}}(-1)^{\nu_C+\nu_D}
\\ \hspace{-3em}\times\,\exp\left[-\lambda\left(T_A+T_B+T_C+T_D\right)/2+\frac{\i}{\hbar}
  \left(S_A+S_C-S_B-S_D\right)\
\right]\nonumber \\\hspace{-3em} \nonumber \times\,\left\langle\!
\exp\!
\left[\frac{\i}{\tau}\left(T_A\{\phi+\textstyle{\frac{e_A}{N}}\}+T_C\{\phi+\textstyle{\frac{e_C}{N}}\}
 -T_B\{\phi-\textstyle{\frac{e_B}{N}}\}-T_D\{\phi-\textstyle{\frac{e_D}{N}}\}\right)\right]\!\right\rangle\,.
\end{eqnarray}
Here  $\mathcal{A},\mathcal{B}$ ($\mathcal{C},\mathcal{D}$)
are the pseudo-orbits associated with the denominator (numerator) of the generating function; $\nu_C$ and $\nu_D$ stand
for the number of orbits  in  $\mathcal{C},\mathcal{D}$.

Next we  turn to the center-phase average over the interval
$-\pi<\phi<\pi$. We write the sawtooth functions as
$\{\phi+\delta\}=\phi+\delta-2\pi\theta(\pi-\phi-\delta)$ and $
\{\phi-\delta\}=\phi-\delta-2\pi\theta(-\pi-\phi+\delta)$, for
$\delta\geq 0$.  The $\phi$ dependence outside the step functions
in the exponent is given by $\i \phi\Delta T/\tau$ where $\Delta
T\equiv T_A+T_C-T_B-T_D$ is the period mismatch between the
pseudo-orbits $\mathcal{A}\cup \mathcal{C}$ and $\mathcal{B}\cup
\mathcal{D}$.  For partner pseudo-orbits the period and action
mismatches are both of order $\hbar$, and therefore $\i \phi\Delta
T/\tau$ can be dropped in the semiclassical limit. The remaining
elementary but cumbersome integral over $\phi$ is periodic in
$\frac{e_{A/B/C/D}}{N}$. The projectors
$\mathcal{P}_-^C\mathcal{P}_-^D$ serve to truncate the respective
Fourier expansions at  $- \frac{N}{2}$; the cumulative periods
$T_{C/D}$ of the pseudo-orbits $\mathcal{C},\mathcal{D}$ are thus
restricted to be smaller than half the Heisenberg time. For
autonomous flows the remaining periodicity in the offset variables
is an artefact of the stroboscopic description. Of physical
interest is only the limit $e_{A/B/C/D}/N\to 0$. In that limit,
the step functions in the exponent can be dropped since they make
only for $\mathcal{O}(\frac{1}{N})$ corrections. The projectors
$\mathcal{P}_-^C\mathcal{P}_-^D$ lose their meaning and are to be
removed, but the soft cutoff for the periods $T_{C/D}$ at half the
Heisenberg time must be kept.  The final result thus reached,
\begin{eqnarray}
  \label{Z1flow3}\nonumber \hspace{-3em}
 Z_{-+}\sim\e^{\i (e_A+e_B-e_C-e_D)/2}
\sum_{\mathcal{A},\mathcal{B},\mathcal{C},\mathcal{D}}^{T_C,T_D<T_H/2}
(-1)^{\nu_C+\nu_D}\, \e^{-\lambda(T_A+T_B+T_C+T_D)/2}\\
\hspace{-1.0em} \times\,
\e^{\i(T_Ae_A+T_Be_B+T_Ce_C+T_De_D)/T_H}\,\e^{\i(S_A+S_C-S_B-S_D)/\hbar}\,,
\end{eqnarray}
is identical with the one previously obtained in
\cite{Heusl07,Muell09,Haake10}.
 In quite analogous
a manner we recover, in our present context, the previous result
$Z_{--}\sim 0$. Equally ignorable is then the distinction of the
zeroth Fourier component $\zeta_0$ and we can simply write
$Z=Z_{-+}+Z_{+-}$.

We would like to note that the reasoning in the present subsection is
not restricted to the unitary symmetry class.

\section{Summary and discussion}\label{summary}

We employ the rigorous Riemann-Siegel lookalike available for
unitary $N\times N$ matrices to treat spectral fluctuations both for
periodically driven and autonomous dynamics.

For periodically driven systems, a finite Gutzwiller trace formula
arises for the traces of powers of the Floquet matrix, $t_n={\rm
  Tr}\,F^n$. Only period-$n$ orbits are involved in $t_n$. The
Riemann-Siegel lookalike then represents the spectral determinant
by pseudo-orbits with cumulative periods up to $n=N/2$, half the
dimensionless Heisenberg time $N$. --- Restricting ourselves to
the unitary symmetry class (no time reversal invariance) we find
the diagonal approximation to yield the exact generating function
of the two-point correlator of the density of quasi-energies
provided by the circular unitary ensemble of random matrices
\cite{Conre07}, for finite matrix dimension $N$. That generating
function reduces to the one for the Gaussian unitary ensemble in
the limit $N\to\infty$. All corrections, due to partnerships of
periodic (pseudo-)orbits generated in close self-encounters, must
therefore cancel. Using computer based summation of the
contributing partner pseudo-orbits we ascertain order-by-order
cancelation in the lowest six orders; the order is given by the
difference $L-V$, with $V$ the overall number of relevant
self-encounters and $L$ their overall number of encounter
stretches.

For autonomous flows, we define a suitable strobe period and the
pertinent time evolution operator to get a Floquet matrix
capturing a finite number $N$ of energy levels as quasi-energies.
The pertinent trace formula, derived from Gutzwiller's one for the
density of energy levels, needs regularization. Again, for the
unitary symmetry class the diagonal approximation yields the exact
finite-$N$ CUE generating function and, with $N\to\infty$, the
exact GUE result. We recover the previously obtained pseudo-orbit
expansions \cite{Heusl07,Muell09,Braun12} for higher-order
corrections in the limit $N\to\infty$, for all Wigner-Dyson
symmetry classes.

A certain charm and, in fact, progress over the previous treatment of
autonomous flows can be seen in two facts. First, the polynomial
character of the secular determinant $\det(1-\e^{\i\varphi}F)$ carries
over to the generating function; the latter retains its polynomial
nature under the pseudo-orbit expansion, even when the restriction
imposed by the Riemann-Siegel lookalike, $n\leq N/2$, is lifted.
Second, only infinitesimal imaginary parts are needed for the
quasi-energy variables. That progress is owed to the rigorous
Riemann-Siegel lookalike.

We would like to pay respect to related work. The explicit
solution of Newton's relations giving the coefficients $A_n$ of
the secular polynomial in terms of the traces $t_n$ has a long
history. Some recent references are
\cite{Krish95,Kotto99,Haake10}. The rigorous finite-$N$
Riemann-Siegel lookalike has been noted before
\cite{Kotto99,Haake10,Berry90,Berry92,Keati92,Bogom92,Eckha00}.
Already more than a decade ago, the possibility of semiclassically
accessing spectral fluctuations using Newton's relations and
Riemann-Siegel for unitary matrices was suggested
\cite{Haake10,Eckha00}. Reasons for expecting cancelation of
contributions from pseudo-orbits with periods beyond half the
Heisenberg time in correlators of spectral determinants were put
forth in \cite{Keati07}. More recently, such cancelation has been
demonstrated for spectral fluctuations in graphs \cite{Band12}.

Postponed to future work is a demonstration of the cancelation of
corrections to all orders of the 'diagrammatic' expansion for
Floquet matrices of the unitary symmetry class. An adaption of the
strategy of \cite{Muell09} should not be overly difficult to come
by, even for the orthogonal and symplectic symmetry classes.
\vspace{0.5cm}

\section*{Acknowledgements}
Support by the Sonderforschungsbereich TR12 'Symmetries and
Universality in Mesoscopic Systems' of the Deutsche
Forschungsgemeinschaft is gratefully acknowledged.

\appendix

\section{Two-point correlator}
\label{2deriv}

The quasi-energy level density can be expressed using a $2\pi$-periodic delta
function, $\rho(\varphi)=\sum_{\mu=1}^{N}\delta(\varphi-\varphi_{\mu}%
)=(2\pi)^{-1}\sum_{n=-\infty}^{\infty}\sum_{\mu=1}^{N} \mathrm{e}^{\i
n(\varphi-\varphi_{\mu}) - |n|0^{+}}$ with real $\varphi$. The factor
$\mathrm{e}^{- |n|0^{+}}$ fattens the delta function to an ordinary function,
though one of arbitrarily small width. Subtracting the contribution $\frac
{N}{2\pi}=\overline\rho$ of $n=0$ we have the fluctuating part of the level
density
\begin{eqnarray}
\label{density}\Delta\rho(\varphi)=\frac{1}{2\pi}\sum_{n=1}^{\infty}%
\mathrm{e}^{- n0^{+}} \big(t_{n}\mathrm{e}^{\i n\varphi}+t_{n}^{*}%
\mathrm{e}^{-\i n\varphi}\big)\,.
\end{eqnarray}

We define the two-point correlator as $\langle\!\langle\Delta
\rho(\varphi)\Delta\rho(\varphi^{\prime})\rangle\!\rangle$ where the
angular brackets demand a double average defined as follows. The
variables $\varphi ,\varphi^{\prime}$ cover a square of area
$(2\pi)^{2}$ in a $\varphi \varphi^{\prime}$-plane. In that square we
have $N^{2}$ pairs of eigenphases $\varphi_{\mu},\varphi_{\nu}$, such
that an area $(\frac{2\pi}{N})^{2}$ can be associated with each
eigenphase pair. The average aimed at must produce a smooth function
of the phase difference $\varphi-\varphi^{\prime}$. We introduce a
real center phase $\phi$ and an offset $e$ as
$\varphi=\phi+\frac{e}{N}$ and $\varphi^{\prime}=\phi-\frac{e
}{N}$. Both new variables range in $2\pi$-intervals, $0\leq\phi<2\pi$
and $-\pi\leq\frac{e}{N}<\pi$. For the double average we choose an
area of the form of a narrow rectangle in the
($\phi\frac{e}{N}$)-plane, with widths $2\pi$ for the center phase
$\phi$ and some $\Delta e/N$ for the offset.  That rectangle must
contain many pairs $\varphi_{\mu},\varphi_{\nu}$, and therefore we
must require $\frac{\Delta e}{N}\gg\frac{2\pi}{N^{2}}$.  That latter
condition is a mild one, allowing the window $\frac{\Delta e}{N}$ to
be much
smaller than the average level spacing.

Now, the average over the center phase $\langle\cdot\rangle=(2\pi)^{-1}%
\int_{0}^{2\pi}d\phi(\cdot)$ can be done and yields, with the help
of the Fourier decomposition (\ref{density}),
\begin{eqnarray}
\label{eq:8}R(e)=\frac{2}{N^{2}}\mathrm{Re}\,\Big\langle\sum
_{n=1}^{\infty}|t_{n}|^{2}\mathrm{e}^{\i2 ne/N-
n0^{+}}\Big\rangle\,.
\end{eqnarray}
The smoothing over the small offset window remains to be done.

It is easy to ascertain that  $R(e)$ equals the real part of
the complex two-point correlator of the resolvent
\begin{equation}
\label{eq:10}C(e)=\frac{2}{N^{2}}\langle\!\langle t\left(\phi+\frac{e}{N}\right)
t\left(\phi-\frac{e}{N}\right)\rangle\!\rangle
\end{equation}
where without loss of generality the offset variable can be taken
with an arbitrary positive imaginary part. Inserting the geometric
series for the resolvent and doing the center-phase average we get
\begin{equation}
\label{eq:12}C(e)=\frac{2}{N^{2}}\Big\langle\sum_{n=1}^{\infty}|t_{n}%
|^{2}\mathrm{e}^{2\i n e/N}\Big\rangle\,.
\end{equation}
Comparison with (\ref{eq:8}) shows that indeed $R(e)=\mathrm{Re}\,C(e)$. The complex correlator,
in turn, is obtained from the generating function as
$\partial_{e_{A}}\partial_{e_{B} }Z\Big|_{e_{A/B/C/D}= e}=
-\frac{1}{4}-\frac{1}{N^{2}} \Big\langle\sum
_{n=1}^{\infty}\mathrm{e}^{\i2en/N}|t_{n}|^{2}\Big\rangle$.

\section{Diagonal approximation: action of projectors
}\label{projector}

For the purposes of the present appendix it is convenient to shorthand as
\[\fl
a=\e^{\i e_{A}/N},\;\;b=\e^{\i e_{B}/N},\;\;c=\e^{\i e_{C}/N},\;\;d=\e^{\i e_{D}/N},\;\;
\delta  =cd,\;\;  F   =\frac{\left(  ab\right)  ^{N/2}}{1-ab}\,.
\]
The building block $\left( \ldots\right)$ in (\ref{(...)})  can then
be written
\begin{eqnarray}
\left(  \ldots\right)    & =F\delta^{-N/2}\frac{\left(
1-ad\right)  \left(
1-bc\right)  }{\left(  1-\delta\right)  }\nonumber\\
& =F\delta^{-N/2}\left[  1-(ad+bc)+ab\delta\right] \sum_{\nu
=0}^{\infty}\delta^{\nu}.
\end{eqnarray}

We now write the components of the generating function obtained by
action of the projectors $\mathcal{P}_-,\mathcal{P}_0$ on $\left(
\ldots\right)$, namely,
\begin{eqnarray}\fl
Z_{-+}^{\mathrm{diag}}
=\mathcal{P}_{-}^{C}\mathcal{P}_{-}^{D}\left(
\ldots\right) \\
\fl\hspace{9mm} =F\left[
\left(\sum_{\nu=-N/2}^{-1}\delta^{\nu}\right)-\left( ad+bc\right)
\left(\sum
_{\nu=-N/2}^{-2}\delta^{\nu}\right)+ab\left(\sum_{\nu=-\left(  N/2-1\right)  }^{-1}%
\delta^{\nu}\right)\right]\,  \nonumber ,
\end{eqnarray}
and
\begin{eqnarray}
Z_{-0}^{\mathrm{diag}}
=\mathcal{P}_{-}^{C}\mathcal{P}_{0}^{D}\left(
\ldots\right)  =-F\frac{a}{c}\,,\nonumber\\
Z_{0+}^{\mathrm{diag}}
=\mathcal{P}_{0}^{C}\mathcal{P}_{-}^{D}\left( \ldots\right)
=-F\frac{b}{d}\,.
\end{eqnarray}
The symmetries (\ref{Zsymmetries}) involve the substitutions $\left(
  c,d,\delta\right) \to \left(1/d, 1/c,1/\delta\right)$ and provide
three more terms of the generating function,
\begin{eqnarray*}
Z_{+-}^{\mathrm{diag}}  & =F\left[
\sum_{\nu=1}^{N/2}\delta^{\nu}-\left(
\frac{a}{c}+\frac{b}{d}\right)
\sum_{\nu=2}^{N/2}\delta^{\nu}+ab\sum_{\nu
=1}^{N/2-1}\delta^{\nu}\right] \, ,\\
Z_{0-}^{\mathrm{diag}}  & =-Fad\,,\\
Z_{+0}^{\mathrm{diag}}  & =-Fbc\,.
\end{eqnarray*}
The last remaining term without a Riemann-Siegel partner is
\[
Z_{00}^{\mathrm{diag}}=\mathcal{P}_{0}^{C}\mathcal{P}_{0}^{D}\left(
\ldots\right)  =F\left(  1+ab\right)  .
\]
Collecting all these terms we obtain,%
\begin{eqnarray*}
Z^{\mathrm{diag}}  & =F\left[
\sum_{\nu=-N/2}^{N/2}\delta^{\nu}-\left( ad+bc\right)
\sum_{\nu=-N/2}^{N/2-1}\delta^{\nu}+ab\sum_{\nu=-\left(
N/2-1\right)  }^{N/2-1}\delta^{\nu}\right] \nonumber \\
& =F\delta^{-N/2}\left[  \frac{1-\delta^{N+1}}{1-\delta}-\left(
ad+bc\right)
\frac{1-\delta^{N}}{1-\delta}+ab\delta\,\frac{1-\delta^{N-1}}{1-\delta
}\right] \, .
\end{eqnarray*}
This expression is equal to
\begin{eqnarray*}
Z^{\mathrm{diag}}  & =F\left[  \delta^{-N/2}\frac{\left(
1-ad\right)  \left( 1-bc\right)  }{\left(  1-\delta\right)
}+\delta^{N/2}\frac{\left(
1-a/c\right)  \left(  1-b/d\right)  }{\left(  1-1/\delta\right)  }\right]  \\
& =\left(  \ldots\right)  +\left.  \left(  \ldots\right)
\right\vert _{\left(  e_{C},e_{D}\right)  \rightarrow\left(
-e_{D},-e_{C}\right)  }
\end{eqnarray*}
which proves Eq. (\ref{Zirni}).

\section*{References}

\bibliographystyle{unsrt}
\bibliography{../pbraun5a}

\begin{thebibliography}{10}

\bibitem{Heusl07}
S.~Heusler, S.~M\"uller, A.~Altland, P.~Braun, and F.~Haake.
\newblock Periodic-orbit theory of level correlations.
\newblock {\em Phys. Rev. Lett.}, 98(4):044103, 2007.

\bibitem{Muell09}
S.~M\"uller, S.~Heusler, A.~Altland, P.~Braun, and P.~Haake.
\newblock Periodic-orbit theory of universal level correlations in quantum
  chaos.
\newblock {\em New Journal of Physics}, 11(10):103025, 2009.

\bibitem{Braun12}
P.~Braun.
\newblock Beyond the {Heisenberg} time: semiclassical treatment of spectral
  correlations in chaotic systems with spin 1/2.
\newblock {\em Journal of Physics A: Mathematical and Theoretical},
  45(4):045102, 2012.

\bibitem{Krish95}
H.~K. Krishnapriyan.
\newblock On evaluating the characteristic polynomial through symmetric
  functions.
\newblock {\em Journal of Chemical Information and Computer Sciences},
  35(2):196--198, 1995.

\bibitem{Kotto99}
T.~Kottos and U.~Smilansky.
\newblock Periodic orbit theory and spectral statistics for quantum graphs.
\newblock {\em Annals of Physics}, 274(1):76 -- 124, 1999.

\bibitem{Haake10}
F.~Haake.
\newblock {\em {Quantum Signatures of Chaos}}.
\newblock Springer Series in Synergetics. Springer, 2010, 2000, 1991.

\bibitem{Berry90}
M.~V. Berry and J.~P. Keating.
\newblock A rule for quantizing chaos?
\newblock {\em Journal of Physics A: Mathematical and General}, 23(21):4839,
  1990.

\bibitem{Berry92}
M.~V. Berry and J.~P. Keating.
\newblock A new asymptotic representation for $\zeta $($\frac{1}{2}$+it) and
  quantum spectral determinants.
\newblock {\em Proceedings of the Royal Society of London. Series A:
  Mathematical and Physical Sciences}, 437(1899):151--173, 1992.

\bibitem{Keati92}
J.~P. Keating.
\newblock Periodic orbit resummation and the quantization of chaos.
\newblock {\em Proceedings of the Royal Society of London. Series A:
  Mathematical and Physical Sciences}, 436(1896):99--108, 1992.

\bibitem{Bogom92}
E.~B. Bogomolny.
\newblock Semiclassical quantization of multidimensional systems.
\newblock {\em Nonlinearity}, 5(4):805, 1992.

\bibitem{Margu04}
G.A. Margulis and R.~Sharp.
\newblock {\em On Some Aspects of the Theory of Anosov Systems}.
\newblock Springer Monographs in Mathematics. Springer, 2004.

\bibitem{Bowen72}
R.~Bowen.
\newblock The equidistribution of closed geodesics.
\newblock {\em Am. J. Math}, 94:413--423, 1972.

\bibitem{Hanna84}
J.~H. Hannay and A.~M. Ozorio De~Almeida.
\newblock Periodic orbits and a correlation function for the semiclassical
  density of states.
\newblock {\em Journal of Physics A: Mathematical and General}, 17(18):3429,
  1984.

\bibitem{Conre07}
J.~B. Conrey, D.~W. Farmer, and M.~R. Zirnbauer.
\newblock Howe pairs, supersymmetry, and ratios of random characteristic
  polynomials for the unitary groups u(n).
\newblock {\em pre-print}, 2007.
\newblock arXiv:math-ph/0511024v2.

\bibitem{Gutzw90}
M.~C. Gutzwiller.
\newblock {\em Chaos in Classical and Quantum Mechanics}.
\newblock Interdisciplinary Applied Mathematics. Springer-Verlag, 1990.

\bibitem{Eckha00}
B.~Eckhardt and U.~Smilansky.
\newblock Stroboscopic quantization of autonomous systems.
\newblock {\em Foundations of Physics}, 31:543--556, 2001.

\bibitem{Keati07}
J.~P. Keating and S.~M\"uller.
\newblock Resummation and the semiclassical theory of spectral statistics.
\newblock {\em Proceedings of the Royal Society A: Mathematical, Physical and
  Engineering Science}, 463(2088):3241--3250, 2007.

\bibitem{Band12}
R.~Band, J.~M. Harrison, and C.~H. Joyner.
\newblock Finite pseudo orbit expansions for spectral quantities of quantum
  graphs.
\newblock {\em pre-print}, 2012.
\newblock arXiv:math-ph/1205.4214.

\end{thebibliography}

\end{document}